\documentclass[preprint]{iacrtrans}
\usepackage[utf8]{inputenc}

\author{Niek J. Bouman} 
\institute{Technische Universiteit Eindhoven, the Netherlands, \email{n.j.bouman@tue.nl}}
\title{Multiprecision Arithmetic for Cryptology in C++}
\subtitle{Compile-Time Computations and Beating the Performance of Hand-Optimized Assembly at Run-Time}

\usepackage{xspace}

\usepackage[frozencache]{minted}
\usepackage{booktabs}

\newcommand{\reftab}[1]{Table~\ref{tab:#1}\xspace}
\newcommand{\refapp}[1]{Appendix~\ref{app:#1}\xspace}
\newcommand{\refsec}[1]{Section~\ref{sec:#1}\xspace}

\newcommand{\ntl}{\textsc{ntl}\xspace}
\newcommand{\gmp}{\textsc{gmp}\xspace}
\newcommand{\libff}{\textsc{libff}\xspace}
\newcommand{\llvm}{\textsc{llvm}\xspace}

\begin{document}

\bibliographystyle{alpha}
\maketitle


\keywords[multiprecision arithmetic, Montgomery representation, Barrett reduction, constant-time algorithms, cryptography, software library, compile-time computations, template metaprogramming]{multiprecision arithmetic \and Montgomery representation \and Barrett reduction \and constant-time algorithms \and cryptography \and software library \and compile-time computations \and template metaprogramming}

\begin{abstract}
  We describe a new C++ library \cite{ctbignum} for multiprecision arithmetic for numbers in the order of 100--500 bits,
i.e., representable with just a few limbs.
The library is written in ``optimizing-compiler-friendly'' C++, with an emphasis on the use of fixed-size arrays and particular function-argument-passing styles (including the avoidance of naked pointers) to allow the limbs to be allocated on the stack or even in registers. Depending on the particular functionality, we get close to, or significantly beat the performance of existing libraries for multiprecision arithmetic that employ hand-optimized assembly code.

Most functions in the library are constant-time, which is a necessity for secure implementations of cryptographic protocols.

  Beyond the favorable runtime performance, our library is, to the best of the author's knowledge, the first library that offers big-integer computations \emph{during compile-time}. 
  For example, when implementing finite-field arithmetic with a fixed modulus, this feature enables the automatic precomputation (at compile time) of the special modulus-dependent constants required for Barrett and Montgomery reduction. Another application is to parse (at compile-time) a base-$10$-encoded big-integer literal.
\end{abstract}
\section{Introduction}
Nowadays, several software libraries for multiprecision arithmetic exist.
One of the most well-known is the \textsl{GNU Multi-Precision library}, or \gmp \cite{granlund}.
Higher-level libraries for multiprecision arithmetic and number theory, like Victor Shoup's \textsl{Number Theory Library} (\ntl) \cite{shoupntl} rely on \gmp for high-performance basic multiprecision arithmetic.

Many modern cryptographic applications, like elliptic-curve-based schemes, blockchains, and multiparty computation frameworks, require multiprecision arithmetic with numbers whose bit-lengths are in the order of 100--500 bits. 
For this bit-range, the use of multiprecision-arithmetical functions designed for arbitrary bit-lengths (meaning that storage must be  allocated dynamically) is typically overkill and could incur a significant runtime overhead; fixed-size arrays seem the better choice. For fixed-bitsize multiprecision numbers, \gmp offers low-level functions whose names are prefixed with ``\texttt{mpn\_}''. Yet other multiprecision-arithmetic libraries wrap those low-level \gmp functions into a more convenient interface and augment them with additional functionality. An example is the \libff  finite-field library from Virza et al. \cite{libff}, which is used in the Zcash implementation of the Zerocash protocol.

%
%
The code base of \gmp consists of a reference implementation in C, combined with many parts written in hand-optimized assembly for specific \textsc{cpu} targets. Also portions of \libff are written in  hand-optimized assembly code.
Given the ongoing development of high-level systems programming languages (e.g., Modern C++, D, Rust) as well as optimizing compilers (e.g., \llvm \cite{lattner2004llvm}), our hypothesis is that replacing code at a high level of abstraction (like C++ code) 
by inline hand-optimized assembly is conceptually wrong:
one loses portability, the code becomes harder to maintain, 
one does not benefit from future compiler improvements, and, if one targets multiple platforms, each platform requires a specific implementation which makes the code base more error-prone. 
This hypothesis is supported by some earlier works, 
for example, \cite{glas, lapackcpp}. 

In this work, we investigate how close we can get to the run-time performance of existing hand-optimized assembly implementations by means of implementing a library of basic routines for multiprecision arithmetic \cite{ctbignum} in Standard C++17, at the time of writing the most recent C++ language dialect. In some sense, our implementations might be viewed as an attempt to 
write ``optimizing-compiler-friendly'' C++ code. 
Note that the choice for C++ is not crucial here, the same techniques can be applied in any high-level compiled language 
with compile-time programming features (such as D). The choice for C++ mainly stems from the author's experience in this language and its wide-spread use.

An important aspect of our design is the frequent use of \texttt{std::array}, the fixed-size array type (introduced in C++11) from the Standard Library. 
We use \texttt{std::array} to pass multiprecision integers into functions, which gives the compiler full knowledge about the length of the multiprecision integer, hence enabling more opportunities for optimization. 

Our library is header-only, mostly because of our frequent use of C++ templates.
An advantage of the header-only property is that the compiler can inline function bodies of our library into client code (which can result in significant performance gains), whereas for separately compiled libraries (such as \ntl)
inlining cannot occur without enabling \emph{link-time optimization} (LTO). 
Note that one drawback of using a library with lots of templated code is an increase in compilation times.


Our experiments confirm that our argument-passing style gives a performance improvement over \gmp's C-style interface, where a multiprecision integer is passed into a function by means of a pointer to memory and a separate integer that contains the length (i.e., the number of limbs). In the latter case, the compiler remains unaware of the relation between the pointer and the (separately specified) length.

Note that in this paper we do not focus on achieving speedups 
by attempting to improve the algebraic complexity of multiprecision arithmetic;
such efforts can be viewed as being ``orthogonal'' to (and would still benefit from) the techniques presented here. 

We highlight several aspects where our library achieves an improvement in terms of runtime performance or usability over state-of-the-art libraries for multiprecision and finite-ring/finite-field arithmetic.
We want to emphasize that 
our library should not be viewed as a full replacement for, say, \gmp, \ntl or \libff, because our library lacks many of the functionalities that such (more mature) libraries provide. (\refapp{overview} gives a high-level overview of the current functionality in our library.)

%

\paragraph{Constant-Time Operations.}
A necessary requirement for the security of a software implementation is that the algorithms should have the \emph{constant time} property, i.e., the execution time of the algorithm may not depend on secret information (such as bits of the cryptographic key), and any branching in the algorithm may not depend on secret information.
Constant-timeness was not a design goal in most libraries for multiprecision arithmetic.
In our library, we have paid special care to ensure that most functions enjoy the constant-time property.

\paragraph{Compile-Time Computations.}
\textsl{Constant expressions} are a notable feature of Modern C++ (introduced in C++11, but extended in more recent versions of the language). Constant expressions, marked by the \texttt{constexpr} keyword, are expressions that can be evaluated at run time as well as at compile time. 
We have paid special attention to ensuring that all of our implementations of the basic algorithms for multiprecision arithmetic are valid \texttt{constexpr} functions.
By doing so (and with the help of some Template Metaprogramming \cite{veldhuizen1996using}), we have created the first library for \emph{compile-time multiprecision arithmetic}. Such compile-time computations can be convenient for specifying pre-computations 
along with the rest of the code, without paying any runtime overhead for these pre-computations.
%
For example, suppose that 
in some function we want to hard-code a multiprecision integer, represented as a human-readable (base-10) digit string. 
\ntl, for example, will parse such a string into a multiprecision integer at runtime, and this happens every time the function is executed, unless the programmer takes explicit care of caching the parsed number. 
With our library, the string representation of the big integer is parsed at compile time, hence incurs zero runtime overhead.
Another example is arithmetic modulo some integer $q$, where the modulus $q$ is constant throughout the program. The reduction modulo $q$ is typically a computationally expensive operation, for which special methods have been devised, like Barrett reduction or the use of Montgomery representation \cite{Menezes1996handbook}.
Our library directly supports to ``hard-code'' this modulus into the Barrett reduction or Montgomery reduction procedures, and to perform the required modulus-dependent pre-computation at compile-time.  



\section{``Compiler-Friendly'' Function-Parameter Passing}
An important aspect of our work is the design of the function prototypes, in other words, the way in which the arguments are passed into, and out of, a function.

Let us first illustrate how a state-of-the-art library for multiprecision arithmetic, \gmp, passes multiprecision integers into functions.
In \gmp, the skeleton of the  \texttt{mpn\_add\_n} function, for adding equi-sized multiprecision integers looks as follows:
\begin{minted}{c}
mp_limb_t
mpn_add_n (mp_ptr rp, mp_srcptr up, mp_srcptr vp, mp_size_t n)
{
  mp_limb_t cy;
  // declaration and initialization of other variables and various asserts 

  do {
      // loop body
  } while (--n != 0);

  return cy; // the last carry is returned
}
\end{minted}
The names \texttt{mp\_ptr} and \texttt{mp\_srcptr} represent pointer types, hence, the function takes three pointers, pointing to the beginning of the memory of the first and second operand and of the result, and it takes the length \texttt{n}. Note that \texttt{mp\_limb\_t} represents the data type of the limbs.

In our library, the function for adding equi-sized multiprecision integers looks as follows.
\begin{minted}{cpp}
template <typename T, size_t N>
constexpr auto add(big_int<N, T> a, big_int<N, T> b) {
  // in this context, 'auto' instructs the compiler to deduce
  // the return type automatically from the return statement

  T carry{};
  big_int<N + 1, T> r{};

  for (auto i = 0; i < N; ++i) {
    // loop body
  }

  r[N] = carry;
  return r;
}
\end{minted}
The \texttt{big\_int} type is a thin wrapper around \texttt{std::array}; a \texttt{big\_int<N, T>} type may safely be read as a \texttt{std::array<T, N>} type. (The exact definition of \texttt{big\_int<N, T>} can be found in \refapp{bigint}.) 
Note that by using this function prototype, the compiler knows exactly the sizes of the inputs. The arguments are passed by value, and the return type, \texttt{big\_int<N + 1, T>}, is returned by value as well. Although the use of pass-by-value might give the impression that many unnecessary copies will be made, from experiments we have observed that a modern optimizing compiler removes/elides unnecessary copies (and in some cases, C++17 actually guarantees copy elision).

\subsection{Representing Compile-Time Arguments}
\label{sec:reprct}
To represent multiprecision integers that are guaranteed to be known at compile time, we use 
\textsl{parameter packs} (a C++11 feature) which give rise to \textsl{variadic templates} (a template class or template function with a variable number of template parameters). To conveniently pass those parameter packs as ``ordinary'' function arguments, we leverage the \texttt{std::integer\_sequence} type, which was introduced in C++14.

An example of the use of \texttt{std::integer\_sequence} is in our function that implements Barrett reduction:
\begin{minted}{cpp}
template <typename T, size_t N, T... Modulus>
constexpr auto 
barrett_reduction(big_int<N, T> x, std::integer_sequence<T, Modulus...>);
\end{minted}
Here, note that \texttt{Modulus} is a variadic template, as indicated by the ellipsis, but can be passed as the second function argument.

\subsection{Enforcing Compile-Time Execution of \texttt{constexpr} Functions}
We also use the \texttt{std::integer\_sequence} type for another, yet related, purpose. 
Given any \texttt{constexpr} function, and constant-expression arguments to that function, the compiler \emph{may} decide to execute that function  at compile time, but is not forced to do so.
Sometimes, we want to enforce the compiler to perform some computation at compile time. 
One ``trick'' to achieve this is to let that \texttt{constexpr}-function produce a result of type \texttt{std::integer\_sequence}, which is then used at another place as input argument. 

\section{``User-Friendlier'' API Without Runtime Penalty}

\subsection{Easy Initialization from a Literal}
With the C++11 feature \emph{User-Defined Literals}, we can define a custom suffix (we have chosen the suffix \texttt{\_Z}) for defining multiprecision-integer literals in base-10 notation, which gets parsed at compile-time.

\paragraph{Example.}
\begin{minted}{cpp}
auto modulus = 144740111546645463731260070504932198000989141205031_Z;
\end{minted}
The \texttt{\_Z} user-defined literal returns an \texttt{std::integer\_sequence}, which means that this return value is guaranteed to be a constant expression. If desired, the object can then be converted into a \texttt{big\_int} type.\footnote{Whether a \texttt{big\_int} type qualifies as a constant expression depends on the context where it appears in the code. For example, if a \texttt{big\_int} is created in the body of a \texttt{constexpr} function, then it qualifies as such; if it is passed into a \texttt{constexpr} function as a function argument, then it does not qualify as a constant expression inside the \texttt{constexpr} function.}
The implementation of our \texttt{\_Z} literal is given in \refapp{lit}.

\subsection{Fast Modular Reductions with Automatic Compile-Time Precomputations}
In \refsec{reprct} we have already shown the function prototype of our \texttt{barrett\_reduction} function. Let us now show how to invoke it, with a modulus that is known at compile time:
\begin{minted}{cpp}
auto num = to_bigint(54766245287875756986436_Z);
auto reduced = barrett_reduction(number, 1180591620717411303449_Z);
assert(reduced == to_bigint(459030734874837027782_Z);
\end{minted}
Because \texttt{barrett\_reduction} is a \texttt{constexpr} function, the following code is also valid:
\begin{minted}{cpp}
constexpr auto num = to_bigint(54766245287875756986436_Z);
auto reduced = barrett_reduction(number, 1180591620717411303449_Z);
static_assert(reduced == to_bigint(459030734874837027782_Z);
\end{minted}

\paragraph{Performing computations in the finite ring $\mathbb{Z}/q\mathbb{Z}$.}
When working in a finite ring or finite field with a fixed modulus, the latter should be defined once and for all. For this scenario, we have created the \texttt{ZqElement} class, which supports operations in the ring $\mathbb{Z}/q\mathbb{Z}$.

First, we declare the type of the ring by calling the \texttt{Zq} template function, which creates a dummy instance of the \texttt{ZqElement} class, of which we take the type using C++'s \texttt{decltype} keyword:

\begin{minted}{cpp}
using GF101 = decltype(Zq(1267650600228229401496703205653_Z));
// define the type of a 101-bit prime field
\end{minted}

Now, we can create instances of our newly created type, and perform arithmetic using the overloaded operators in the ring:

\begin{minted}{cpp}
GF101 x(8732191096651392800298638976_Z);
GF101 y(27349736_Z);

auto sum = x + y;
auto prod = x * y;
\end{minted}

\subsection{Functional Form vs. Procedural Form}
In \ntl, the programmer can express computations in two forms, the \emph{functional form} and the \emph{procedural form}.
For example, if \texttt{x}, \texttt{a}, and \texttt{b} are variables of \ntl's heap-allocated big-integer type \texttt{ZZ}, the functional form of an addition would look like $\texttt{x = a + b}$, while the procedural variant would look like \texttt{add(x, a, b)}.
While the functional form is (arguably) more natural to write, the  procedural version is recommended for getting the best performance, as the functional form might generate temporary objects.

Our library does not use dynamic memory allocation, which means that the compiler can easily optimize-away temporaries arising from a functional calling style.

\subsection{Composition with Higher-Level Libraries}
Because our \texttt{big\_int} type is essentially an array, it is easy and safe (i.e., no risk of memory leaks) to use the type inside some other type, like a standard library container, such as \texttt{std::vector}.

\paragraph{Example: Matrix Multiplication via the Eigen library.}
Suppose that we need to use matrices, and the associated operations, like matrix multiplication.
\ntl, for example, defines its own dedicated vector and matrix class.
As an alternative, we would like to compose our \texttt{big\_int} or \texttt{Zq\_Element} type with an existing matrix class.
As a concrete working example, let us consider \emph{Eigen} \cite{eigenweb}, a C++ library for linear algebra that offers a templated Matrix class whose element type can be specified as a template parameter.
Suppose that we would like to define matrix multiplication in the field modulo the $101$-bit prime $p = 1267650600228229401496703205653$.

The first step is to provide Eigen with information about our type by means of a traits class.
We can do this by means of some ``boilerplate'' code, which we show in \refapp{eigentrait}. Then, we can define the matrix type (in this example, a fixed-size $3$-by-$3$ matrix) over our field and perform an actual computation:
\begin{minted}{cpp}
using Mx33 = Eigen::Matrix<GF101, 3, 3>;
Mx33 A, B, C;

A <<  2_Z,   4_Z,  6_Z,
     10_Z,  11_Z, 12_Z,
      1_Z, 100_Z, 30_Z;

B << 5_Z,  3_Z,  9_Z,
     8_Z,  6_Z, 55_Z,
     3_Z, 17_Z,  2_Z;

C = A * B;
\end{minted}

%

\section{Running-Time Benchmarks}
\label{sec:bench}
We have run the benchmarks described below on a MacBook Pro (2017) quad-core 2300 MHz Intel Core i5 7th generation ``Kaby Lake'' processor with 16 GB memory. The ``Turbo Boost'' feature (which is a dynamic overclocking feature on some Intel \textsc{cpu}s) has been explicitly disabled using a small program called ``Turbo Boost Switcher''.
We have used Apple's LLVM compiler Version 9.0.0, with \texttt{-O3} optimizations; we did not enable LTO.
For obtaining the actual timings, we have used Google's Benchmark library \cite{gbench}.

Note that we apply the functions that we benchmark to random inputs \emph{generated at runtime}, to rule out the possibility that the compiler can already compute the result at compile time (in that case, the code that we want to benchmark would disappear from the benchmark's hot loop).

\subsection{Multiplication}
\reftab{mult} shows benchmarks for multiplication.
Based on inspection of the assembly code of the compiled benchmark, we
believe that the 
speedup over \gmp (and \ntl) that we get for small limb sizes ($n \in \{2,3,4\}$) is because the compiler has inlined our multiplication function (hence no function-call overhead) and has fully unrolled the \texttt{for}-loop (because the compiler knows the loop lengths at compile time due to the use of C++ templates).

At some point (beyond $4$ limbs) our implementation loses from \gmp in terms of speed. By compiling with \texttt{-march=native}, see \reftab{mult2}, our implementation becomes a bit faster, especially for $8$ limbs, where our multiplication method is significantly slower than \gmp.

Another reason for the significant slowdown for the $8$-limb case seems to be the fact that the compiler does not inline the call to the multiplication function. By additionally enforcing inlining, we obtain the results in \reftab{mult3}.  Such inlining can be enforced using compiler attributes: 
\texttt{[[gnu::always\_inline]]}.



\begin{table}
  \caption{Time (\textsc{cpu} time) spent to multiply two $n$-limb operands (limb size: $64$-bit). 
  \label{tab:mult}
  Comparison between our implementation of $O(n^2)$-time schoolbook multiplication, \gmp's \texttt{mpn\_mul} and \ntl's \texttt{mul} function.} 
  \centering
  \begin{tabular}{rrrr}
\toprule 
    $n$          &  Our library & \gmp & \ntl \\ 
  \midrule
    2          &  5 ns& 13 ns  & 22 ns  \\
    3          & 11 ns& 22 ns  & 32 ns \\
    4          & 19 ns& 30 ns  & 38 ns \\
    5          & 39 ns& 36 ns  & 47 ns \\
    6          & 56 ns& 45 ns  & 54 ns \\
    7          & 72 ns& 58 ns  & 65 ns\\
    8         & 125 ns& 69 ns  & 77 ns \\
\bottomrule
\end{tabular}
\end{table}

\begin{table}
  \caption{Same experiment as \reftab{mult}, but compiled with \texttt{-march=native} flag.}
  \label{tab:mult2}
  \centering
  \begin{tabular}{rrrr}
\toprule 
    $n$          &  Our library & \gmp & \ntl \\ 
  \midrule
    2          &  4 ns& 14 ns  & 24 ns  \\
    3          & 11 ns& 22 ns  & 34 ns \\
    4          & 18 ns& 29 ns  & 38 ns \\
    5          & 36 ns& 37 ns  & 47 ns \\
    6          & 50 ns& 45 ns  & 58 ns \\
    7          & 67 ns& 56 ns  & 65 ns\\
    8         & 102 ns& 68 ns  & 77 ns \\
\bottomrule
\end{tabular}
\end{table}

\begin{table}
  \caption{Same experiment as \reftab{mult}, but compiled with \texttt{-march=native} flag and forced inlining of our \texttt{mul} function.}
  \label{tab:mult3}
  \centering
  \begin{tabular}{rrrr}
\toprule 
    $n$          &  Our library & \gmp & \ntl \\ 
  \midrule
    2          &  4 ns& 14 ns  & 22 ns  \\
    3          &  9 ns& 22 ns  & 34 ns \\
    4          & 18 ns& 29 ns  & 38 ns \\
    5          & 34 ns& 36 ns  & 46 ns \\
    6          & 49 ns& 45 ns  & 54 ns \\
    7          & 70 ns& 55 ns  & 65 ns\\
    8         &  85 ns& 67 ns  & 78 ns \\
\bottomrule
\end{tabular}
\end{table}

\begin{table}
  \caption{Time (\textsc{cpu} time) spent to perform Montgomery multiplication  two $4$-limb operands (limb size: $64$-bit). 
  \label{tab:montmult}
  Comparison between our ``textbook implementation'' of Montgomery  multiplication (following \cite[Chap.~14]{Menezes1996handbook}), and \libff's \texttt{mul\_reduce} function.} 
  \centering
  \begin{tabular}{cccc}
\toprule 
    $n$          & \multicolumn{2}{c}{Our library}    & \libff \texttt{mul\_reduce} \\ 
             &  (no inlining) & (enforced inlining) &   \\ 
  \midrule
    4          & 87 ns& 39 ns  & 66 ns \\
\bottomrule
\end{tabular}
\end{table}

\subsection{Montgomery Multiplication}
With respect to Montgomery multiplication, we compare our library against \libff.
\reftab{montmult} shows a benchmark for Montgomery multiplication of $4$-limb operands.
The $4$-limb specialization of the \texttt{mul\_reduce} function of \libff is written by hand in assembly, and implements the CIOS method (Coarsely Integrated Operand Scanning) \cite{koc1996analyzing}. From inspecting the compiler-generate assembly code of the benchmark program, we have observed that \libff's Montgomery multiplication code gets inlined.  
We beat the performance of \libff by forcing the compiler to also inline our 
\texttt{montgomery\_mul} code. 
If we do not enforce inlining, then \libff's (inlined) assembly code is faster, which seems to indicate that in this case the gap in running time is merely due to function-call overhead.

In other words, it seems that the use of hand-written inline assembly in \libff triggered the compiler to inline the multiplication function in our benchmark program, and that this inlining operation achieves a performance lead over \ntl; \libff's hand-written assembly language for the $4$-limb case in the multiplication function (that is the one that we benchmarked) actually seems to be slower than what the compiler generates out of our C++ implementation. 
\subsection{Modular Exponentiation}
\reftab{modexp} shows benchmarking results for modular exponentiation.
Given a fixed $200$-bit modulus, we raise a $195$ bit operand to an exponent of $122$ bits.
Again, we observed that applying forced inlining of \texttt{montgomery\_mul} (a subroutine of our modular-exponentiation function \texttt{mod\_exp}) has a big impact on the performance.
\begin{table}
  \label{tab:modexp}
  \caption{Comparison of modular exponentiation: given a modulus of $200$ bits, we raise a $195$ bit operand to an exponent of $122$ bits.
  We compare our ``textbook implementation'' of Montgomery-multiplication-based modular exponentiation (following \cite[Chap.~14]{Menezes1996handbook}), to \ntl's \texttt{power} function and \libff's power operator. } 
  \centering
  \begin{tabular}{cccc}
\toprule 
            \multicolumn{2}{c}{Our library} & \libff  &   \ntl  \\ 

               (no inlining) & (enforced inlining) &   \\ 
  \midrule
           15474 ns &  6594 ns & 10530 ns   & 16862 ns \\

\bottomrule
\end{tabular}
\end{table}

\section{Conclusion}
The running-time benchmarks provide some experimental evidence for our hypothesis that, with respect to multiprecision integer arithmetic, the combination of high-level code and modern compilers has become a serious alternative for hand-optimized assembly code. 
Furthermore, with modern high-level systems languages, e.g. modern  C++, we can create easy-to-use APIs without paying in terms of runtime overhead.  

The functions in our library enable multi-precision and modular arithmetic at \emph{compile-time}, which is useful when all inputs to some computation are known at compile-time (and in this case those computations will have zero cost at runtime).
For our targeted small-number-of-limbs scenario, the \emph{run-time} performance of the \emph{same} functions is competitive with or even better than several  existing state-of-the-art libraries for multiprecision arithmetic. The code base of our library is pure C++ and much more concise than, say, the parts of \gmp that implement the same functionality. Moreover, the code parts written in hand-optimized assembly in libraries like \gmp prevent (the currently available) compilers from further optimizing those pieces of code.

We have released our work as an open-source library \cite{ctbignum} under a permissive license (Apache 2) and we invite others to contribute to its development. At the same time, we hope that some of the techniques presented in this work can contribute to or influence the development of the existing software libraries for cryptography and number theory.

Although we used C++ for this work, the ideal programming language for expressing algorithms that compile to optimal code might not yet exist, according to the following quotes from two greats. 

\begin{quote} 
  \emph{``For 15 years or so I have been trying to think of how to write a compiler that really produces top quality code. 
[\ldots]
  We found ourselves always running up against the same problem: the compiler needs to be in a dialog with the programmer; it needs to know properties of the data, and whether certain cases can arise, etc. And we couldn't think of a good language in which to have such a dialog.''}
  --- D. E. Knuth \cite{Knuth1974}
\end{quote}

\begin{quote} 
  \emph{``Make sure that the language is sufficiently expressive to enable
  most other optimizations to be made in the language itself''}
  --- C. A. R. Hoare \cite{hoare1973hints}
\end{quote} 

\begin{quote} 
\emph{``Once we have a suitable language, we will be able to have what seems to be emerging as the programming system of the future: an interactive program-manipulation system [\ldots].
  The programmer using such a system will write his beautifully-structured, but possibly inefficient, program P; then he will interactively specify transformations that make it efficient.''}
--- D. E. Knuth \cite{Knuth1974}
\end{quote}
One might argue, however, that \llvm is actually an instance of Knuth's ``interactive program-manipulation system''.  
To the best of the author's knowledge, there is not yet a language that lets you express optimization hints along with program logic. On the other hand, by mixing code (like C++) with compiler pragmas and attributes (like \texttt{[[gnu::always\_inline]]}), one comes fairly close to what Hoare envisioned.

\section*{Acknowledgment}
The author wishes to thank Berry Schoenmakers and Niels de Vreede for helpful discussions, and Johan Engelen for his help with improving the code base and preparing the repository for an open-source release, and for reviewing this paper.

\bibliography{refs}
\appendix
\section{Overview of the Library Functionality}
\label{app:overview}
\begin{itemize}
\item Basic arithmetic
\begin{itemize}
\item Addition / Subtraction
\item Multiplication (naive $O(n^2)$ ``schoolbook'' algorithm)
\item Division: short division (single-limb divisor) and Knuth's ``algorithm D''
\item Bitwise / Limbwise shifts
\end{itemize}
\item Relational operators
\begin{itemize}
\item Equality
\item Comparison operators
\end{itemize}
\item Modular operations
\begin{itemize}
\item Modular addition,
\item extended GCD and modular inverse (currently: compile-time only),
\item Barrett reduction,
\item Montgomery reduction,
\item Montgomery multiplication,
\item Modular exponentiation (based on Montgomery multiplication)
\end{itemize}
\item I/O
\begin{itemize}
\item  Parsing of base-10 literal
\item Output to base-10 representation 
\end{itemize}
\end{itemize}

\section{Some Source-Code Fragments of Our Library}
\subsection{The \texttt{big\_int} Type}
\label{app:bigint}
\begin{minted}{cpp}
template <size_t N, typename T = uint64_t,
          typename = std::enable_if_t<std::is_integral<T>::value>>
struct big_int {
  std::array<T, N> repr;

  constexpr T &operator[](size_t i) { return repr[i]; }
  constexpr const T &operator[](size_t i) const { return repr[i]; }
  constexpr size_t size() const noexcept { return repr.size(); }
};
\end{minted}
\subsection{Function to compute the corresponding double-bitlength type}
\begin{minted}{cpp}
template <typename T> struct dbl_bitlen { using type = void; };
template <> struct dbl_bitlen<uint8_t> { using type = uint16_t; };
template <> struct dbl_bitlen<uint16_t> { using type = uint32_t; };
template <> struct dbl_bitlen<uint32_t> { using type = uint64_t; };
template <> struct dbl_bitlen<uint64_t> { using type = __uint128_t; };
\end{minted}
\subsection{Addition}
\begin{minted}{cpp}
template <typename T, size_t N>
constexpr auto add(big_int<N, T> a, big_int<N, T> b) {
  T carry{};
  big_int<N + 1, T> r{};

  for (auto i = 0; i < N; ++i) {
    auto aa = a[i];
    auto sum = aa + b[i];
    auto res = sum + carry;
    carry = (sum < aa) | (res < sum);
    r[i] = res;
  }

  r[N] = carry;
  return r;
}
\end{minted}
\noindent
The following function is used in \refapp{lit}.
\begin{minted}{cpp}
template <typename T, size_t N1, size_t N2>
constexpr auto accumulate(big_int<N1, T> accum, big_int<N2, T> b) {
  T carry{};
  big_int<N1, T> r{};
  auto m = std::min(N1, N2);
  for (auto i = 0; i < m; ++i) {
    auto aa = accum[i];
    auto sum = aa + b[i];
    auto res = sum + carry;
    carry = (sum < aa) | (res < sum);
    r[i] = res;
  }
  if (N1 > N2)
    r[N2] = carry;
  return r;
}
\end{minted}
\subsection{Multiplication}
\begin{minted}{cpp}
template <size_t padding_limbs = 0, size_t M, size_t N, typename T>
constexpr auto mul(big_int<M, T> u, big_int<N, T> v) {

  using TT = typename dbl_bitlen<T>::type;
  big_int<M + N + padding_limbs, T> w{};
  for (auto j = 0; j < N; ++j) {
    T k = 0;
    for (auto i = 0; i < M; ++i) {
      TT t = static_cast<TT>(u[i]) * static_cast<TT>(v[j]) + w[i + j] + k;
      w[i + j] = static_cast<T>(t);
      k = t >> std::numeric_limits<T>::digits;
    }
    w[j + M] = k;
  }
  return w;
}
\end{minted}
\noindent
The following function is used in \refapp{lit}.
\begin{minted}{cpp}
template <size_t ResultLength, size_t M, size_t N, typename T>
constexpr auto partial_mul(big_int<M, T> u, big_int<N, T> v) {
  using TT = typename dbl_bitlen<T>::type;
  big_int<ResultLength, T> w{};
  for (auto j = 0; j < N; ++j) {
    T k = 0;
    const auto m = std::min(M, ResultLength - j);
    for (auto i = 0; i < m; ++i) {
      TT t = static_cast<TT>(u[i]) * static_cast<TT>(v[j]) + w[i + j] + k;
      w[i + j] = static_cast<T>(t);
      k = t >> std::numeric_limits<T>::digits;
    }
    if (j + M < ResultLength)
      w[j + M] = k;
  }
  return w;
}
\end{minted}

\subsection{Literal Initialization}
\label{app:lit}
\begin{minted}{cpp}
namespace detail {

template <size_t N, typename T>
constexpr auto tight_length(big_int<N, T> num) {
  // count the effective number of limbs
  // (ignoring zero-limbs at the most-significant-limb side)
  size_t L = N;
  while (L > 0 && num[L - 1] == 0)
    --L;
  return L;
}

template <typename T = uint64_t, char... Chars>
constexpr auto chars_to_big_int(std::integer_sequence<char, Chars...>) {
  // might return a 'non-tight' representation, meaning that there could be
  // leading zero-limbs
  constexpr size_t len = sizeof...(Chars);
  constexpr size_t N = 1 + (10 * len) / (3 * std::numeric_limits<T>::digits);
  std::array<char, len> digits{Chars...};
  big_int<N, T> num{0};
  big_int<N, T> power_of_ten{1};

  for (int i = len - 1; i >= 0; --i) {
    num = accumulate(
        num, partial_mul<N>(big_int<1, T>{static_cast<T>(digits[i]) - 48},
                            power_of_ten));
    power_of_ten =
        partial_mul<N>(big_int<1, T>{static_cast<T>(10)}, power_of_ten);
  }
  return num;
}

template <typename T = uint64_t, char... Chars, size_t... Is>
constexpr auto chars_to_integer_seq(std::integer_sequence<char, Chars...>,
                                    std::index_sequence<Is...>) {
  constexpr auto num =
      detail::chars_to_big_int(std::integer_sequence<char, Chars...>{});
  return std::integer_sequence<T, num[Is]...>{};
}

} // end of detail namespace

template <char... Chars> constexpr auto operator"" _Z() {
  // this function performs a conversion from base-10 to base-2^W twice:
  // first, to compute the tight length L
  // second, to compute the integer sequence

  using T = uint64_t;

  constexpr auto num =
      detail::chars_to_big_int(std::integer_sequence<char, Chars...>{});
  constexpr auto L = detail::tight_length(num);

  return detail::chars_to_integer_seq<T>(
      std::integer_sequence<char, Chars...>{}, 
      std::make_index_sequence<L>{});
}

template <size_t ExplicitLength = 0, typename T, T... Limbs>
constexpr auto to_big_int(std::integer_sequence<T, Limbs...>) {
  return big_int<ExplicitLength ? ExplicitLength : sizeof...(Limbs),T>
  { Limbs... };
}

\end{minted}

\subsection{Eigen NumTraits Example}
\label{app:eigentrait}
\begin{minted}{cpp}
namespace cbn {
using GF101 = decltype(Zq(1267650600228229401496703205653_Z));
}
namespace Eigen {
template <>
struct NumTraits<cbn::GF101> : GenericNumTraits<cbn::GF101>
{
  using Real = cbn::GF101;
  using NonInteger = cbn::GF101;
  using Literal = cbn::GF101;
  using Nested = cbn::GF101;
  static inline Real epsilon() { return 0; }
  static inline Real dummy_precision() { return 0; }
  static inline int digits10() { return 0; }
  enum {
    IsComplex = 0, IsInteger = 1, IsSigned = 1,
    RequireInitialization = 1, ReadCost = 1,
    AddCost = 1, MulCost = 1
  };
};
}
\end{minted}


\end{document}